\documentclass[twocolumn,showpacs,preprintnumbers,amsmath,amssymb,epsfig]{revtex4}

\usepackage{graphicx}
\usepackage{dcolumn}
\usepackage{epsfig}

\def\ga{\mathrel{\mathpalette\fun >}}
\def\fun#1#2{\lower3.6pt\vbox{\baselineskip0pt\lineskip.9pt
  \ialign{$\mathsurround=0pt#1\hfil##\hfil$\crcr#2\crcr\sim\crcr}}}

\input epsf
\def\plotone#1{\centering \leavevmode
\epsfxsize= 1.0\columnwidth \epsfbox{#1}}

\def\be{\begin{equation}}
\def\ee{\end{equation}}
\def\ba{\begin{eqnarray}}
\def\ea{\end{eqnarray}}
\def\half{\frac{1}{2}}
\def\nn{\nonumber}

\begin{document}
\bibliographystyle{prsty}
\preprint{}

\title{The Detectability of Departures from the Inflationary Consistency
Equation}

\author{Yong-Seon Song}
\email{yssong@bubba.ucdavis.edu}
\affiliation{
Department of Physics, One Shields Avenue, University of California, Davis, California 95616
}

\author{Lloyd Knox}
\email{lknox@ucdavis.edu}
\affiliation{
Department of Physics, One Shields Avenue, University of California, Davis, California 95616
}

\date{\today}

\begin{abstract}
We study the detectability, given CMB polarization maps, 
of departures from the inflationary
consistency equation, $r \equiv T/S \simeq -5 n_T$, where $T$ and $S$
are the tensor and scalar contributions to the quadrupole variance,
respectively.  The consistency equation holds if inflation is driven
by a slowly-rolling scalar field.  Departures can be caused by: 1)
higher-order terms in the expansion in slow-roll parameters, 2)
quantum loop corrections or 3) multiple fields.  Higher-order
corrections in the first two slow-roll parameters are undetectably
small.  Loop corrections are detectable if they are nearly maximal and
$r \ga 0.1$.  Large departures $(|\Delta n_T| \ga 0.1)$ can be seen if
$r \ga 0.001$.  High angular resolution can be important for detecting
non-zero $r+5n_T$, even when not important for detecting non-zero $r$.
\end{abstract}

\pacs{draft}


\maketitle

\section{Introduction}

Inflation is the most promising paradigm for explaining our flat, old
and structure-filled Universe.  Recent observations of the CMB
temperature power spectrum \cite{hinshaw03,pearson02,kuo02} confirm
that the scalar perturbation spectrum is nearly scale-invariant, as
predicted.  Further, the detection of the temperature/polarization
E-mode anti-correlation on degree scales \cite{kogut03} is a strong
indication of correlations on length scales larger than the classical
horizon \cite{hu97,spergel97} \footnote{In particular this anti-correlation
rules out the `mimic' models \cite{turok96a} designed to create
inflation-looking temperature power spectra without such long-range
correlations.}.

In addition to the scalar spectrum of perturbations studied so far, inflation
also produces tensor perturbations.  Several exciting possibilities could 
come from study of the 
amplitude and shape of this tensor perturbation
spectrum:  1) single-field slow-roll inflation can be verified through
confirmation of the consistency equation;
2) the presence of loop corrections can be inferred (from small,
but detectable, departures from the consistency equation)
and used to constrain the more fundamental physics underlying
the effective field theory description of the inflaton; or 3) 
single-field slow-roll inflation
can be ruled out if the departures from the consistency equation
are larger than can come from loop corrections.  Here we quantify
how well the tensor spectrum can be measured (for varying sensitivity
and angular resolution of CMB observations) and discuss these
various possibilities.

That there must be a consistency equation can be seen from
a degrees-of-freedom counting argument.  For a single slowly-rolling
scalar field, to leading order in the expansion in slow roll parameters,
there are only three important parameters: the Hubble parameter and its first
two derivatives with respect to the scalar field $\phi$.  
Since these
three parameters control four observables (the amplitude and power-law
indexes of the tensor and scalar perturbation spectra) these
four observables cannot be independent and indeed are related by
$r \simeq -5n_T$.  

The scalar field driving inflation is most likely
the scalar field of an effective field theory (EFT),which may
receive large corrections near some high energy scale, $M$.
The degrees
of freedom at this higher scale can generate quantum loop corrections
to the effective Lagrangian which will
lead to departures from the consistency equation \cite{kaloper02}.  

We show here that these departures may be large enough to be detectable, via
CMB polarization observations,
if $r \ga 0.1$.  Thus there may be an observational window on the
more fundamental physics underlying the inflaton. For this window
to be open, $M^2$ must not be much larger than $H^2$.
Holographic considerations may place an upper
bound on $M$ \cite{albrecht02}.

We further show that the quantum loop corrections envisioned in
\cite{kaloper02} cannot lead to large departures from the
consistency equation.  Thus large departures cannot be confused
with these loop corrections, but would clearly signal the failure of
a single-field description.

We discuss the implications of our results for observation strategies.
Although the high resolution required to reduce the gravitational lensing
contamination of the tensor signal is not necessary for measurement
of the amplitude of the tensor spectrum when $r > 0.1$, it can make a 
significant difference for measurement of the shape.  However,
for $r \simeq 0.01$ high resolution has very little benefit since other
noise sources dominate.

We concentrate solely on CMB observations because these are likely the
only observations that can be used to detect the influence of tensor
perturbations from inflation.  Although direct detection by
space-based interferometers has been discussed \cite{great}, such a
mission is at least several decades away and it is likely that
foreground signals (from merging massive black hole binaries) will
dominate the primordial signal \cite{jaffe03}.  If direct detection were
possible, in combination with CMB observations it would be enormously
valuable for measuring the shape of the tensor spectrum since the
length scales probed differ by 10 orders of magnitude \cite{turner02}.

In section I we introduce the consistency equation to leading order
and its next order corrections. In section II we discuss the 
loop corrections.  In section III we discuss CMB observations and how
the tensor spectrum can be recovered from them.  In section
IV we present our detectability limits for departures from
the consistency equation and the presence of loop corrections.

\section{The consistency equation}
We consider a single scalar field slow-roll inflation model \cite{albrecht82}.
For a review of the spectrum of scalar and tensor perturbations
produced by a slow-roll scalar field, see \cite{lidsey97}.
The equation of motion for the single scalar field 
in an expanding universe is
\be
\ddot{\phi}+3H\dot{\phi}+V'=0,
\ee
where a dot denotes the derivative in terms of a physical time $t$
and a prime denotes the derivative in terms of $\phi$. 
The Friedmann equation with the scalar field is
\be
H^2={8\pi \over m^2_{pl}}\left(\half \dot{\phi}^2 +V(\phi) \right).
\ee
The slow-roll parameters, $\epsilon_0$ and $\eta_0$, which we use here
are defined with the Hubble parameter $H$ rather than potential $V$ \cite{liddle94},
in order to clarify the relation with the loop corrections. 
They are given by
\ba
\epsilon_0 &=& {m^2_{pl} \over 4\pi}\left( {H' \over H}\right)^2 \nn \\
\eta_0 &=& {m^2_{pl} \over 4\pi}{H'' \over H}.
\ea

The square roots of the resulting scalar and tensor power spectra are,
to next order in the slow-roll parameters $\epsilon_0$ and $\eta_0$
\cite{stewart93,abbott84} \ba\label{amp}
A_S^0(k)&=&\frac{2}{5}\left[1-(2C+1)\epsilon_0+2\eta_0\right]
{2 \over m_{pl}^2}\frac{H^2}{|H'|}\,\,\bigg|_{k=aH} \nn \\
A_T^0(k)&=&\frac{1}{10}\left[1-(C+1)\epsilon_0\right]{4
\over \sqrt{\pi}} \frac{H}{m_{pl}}\,\,\bigg|_{k=aH} \ea where $C\simeq
-0.73$.  The ratio between $(A_S^0(k))^2$ and $(A_T^0(k))^2$ is simply
$\epsilon_0$ to zeroth order.  The full description to the next order
is \cite{lidsey97} \be \epsilon_0\simeq{(A_T^0)^2 \over (A_S^0)^2}
\left[ 1-2C(\epsilon_0-\eta_0)\right].  \ee This is the first step
toward getting the consistency equation in a simple algebraic form.
The ratio of both amplitudes turns into a slow-roll parameter
$\epsilon_0$.

The tensor power spectrum is determined by an inflationary energy scale
represented by $H$.
Thus its spectral index only includes the first order derivative to leading 
order. 
The spectral index $n_T^0$ for the tensor power spectrum in the next order is
\cite{liddle92} 
\ba\label{ten1}
n_T^0\simeq -2\epsilon_0 \left[1+(3+2C)\epsilon_0-2(1+C)\eta_0\right].
\ea

Two degrees of freedom of the primordial perturbations, $H$ and $\epsilon_0$,
describe the three observables, $A_S^0$, $A_T^0$ and $n_T$.
There is a single equation which relates these observables.
It can be written as 
\ba
n_T^0 + 2{(A_T^0)^2 \over (A_S^0)^2}\simeq& 0,
\ea
to zeroth order.
If we consider the possible departure in the next order, the consistency equation
is \cite{copeland94}
\ba\label{con1}
n_T^0 + 2{A_T^{0\,2} \over A_S^{0\,2}} - 2{A_T^{0\,4} \over A_S^{0\,4}}
+2{A_T^{0\,2} \over A_S^{0\,2}}\left(1-n^0_S\right)&\simeq& 0.
\ea
The next order corrections can be estimated from the observables,
such as $(A_T^0/A_S^0)^2$ and $n_S$.
Such a departure is not unknown quantity like the loop corrections
which we present in the next section.

We expect higher-order derivatives to be negligible.
If they are not then we could tell from
observing the scalar spectrum
and modify our analysis accordingly.

\section{The loop corrections}
The large corrections to EFT appearing near the higher energy scale $M$
may leave an imprint on the CMB.
They appear in the effective Lagrangian due to 
loop corrections proportional to 
the even powers of $H/M$ where $H<M$ \cite{kaloper02}.
The detectability of such `short distance physics' will be possible
only if $M$ is far less than $M_{Pl}$.
Some theories allow such a low scale of $M$.

The variance of density perturbations $\delta \rho/\rho$ is proportional 
to the mean-square spectrum of fluctuations of the scalar field as
\be
\left({\delta\rho \over \rho}\right)^2 \propto
\left(\frac{H}{\dot{\phi}}\right)^2 (\delta \phi)^2.
\ee
The quantum fluctuation $(\delta \phi)^2$ is 
given by the equal time two point function 
$\langle\phi(p)\phi(-p)\rangle$ when the mode crosses out the horizon.
The physical momentum $p$ is equal to $H$ at the horizon crossing.
Thus we have
\ba
(\delta \phi)^2 \sim \langle\phi(p)\phi(-p)\rangle \mid_{p=H},
\ea
where the two point function is defined as
\ba\label{two}
\langle\phi(p)\phi(-p)\rangle = \half (p^2+H^2),
\ea
in de Sitter space. 

The loop corrections to
to $\langle\phi(p)\phi(-p)\rangle$ are \cite{kaloper02}.
\ba
\langle\phi(p)\phi(-p)\rangle \,\big|_{p=H}\,\,\, =\,\,
H^2\left(1+\chi \frac{H^2}{M^2}+\,\cdot\cdot\cdot\,\right),
\ea
where the curvature is kept close to $H^2$ 
by fine-tuning.

The appearance of the loop corrections in the density perturbations
leads to a shift in the observable quantities 
from the zeroth order, as
\ba\label{loop}
A_S&=&A_S^0\left(1+\chi_S\frac{H^2}{M^2}\right) \nn \\
A_T&=&A_T^0\left(1+\chi_T\frac{H^2}{M^2}\right) \nn \\
n_T&=&n_T^0\left(1+2\chi_T\frac{H^2}{M^2}\right),
\ea
where we separate $\chi$ into $\chi_S$ for the scalar
and  $\chi_T$ for the tensor.  The shift in $A_S$ \footnote{The 
shifts in $n_T$ and $A_T$ cancel eachother.} leads to
an alteration of the consistency equation such that
\ba\label{consis}
n_T+2\left(\frac{A_T}{A_S}\right)^2\simeq-2\epsilon_0\chi_S\frac{H^2}{M^2}.
\ea
Thus an observed violation of the consistency equation is
possibly a signature of 
short distance degrees of freedom affecting the inflaton \cite{kaloper02}.

There are many different ways in which the EFT could receive
large corrections near some mass scale $M$.  It could happen
simply from a Yukawa interaction with a fermion of mass $M$.
Or, if we have the high
dimensional theory with proper compactification, then $M$ is the
reduced Planck scale.  In M theory, the fundamental scale possibly
approaches the scale of $H$ and can give us the detectable loop
corrections, $\chi (H^2/M^2)>0.1$ \cite{kaloper02}.
Holographic considerations also suggest large corrections to EFT at
short distances.

With $|\chi|<1$, we have the maximum departure from
the loop corrections as
\ba
\bigg| \,\,n_T+2\left(\frac{A_T}{A_S}\right)^2\bigg| < 2\epsilon_0,
\ea
since $H^2/M^2$ is always less than unity.
At $|\chi (H^2/M^2)|\sim 1$, the consistency equation is maximally 
broken by the loop corrections.
As we discuss below, this maximal departure is larger than
any other contribution in the single scalar field slow-roll
inflationary model.

\begin{figure}[htbp]
\label{f1}
  \begin{center}
    \plotone{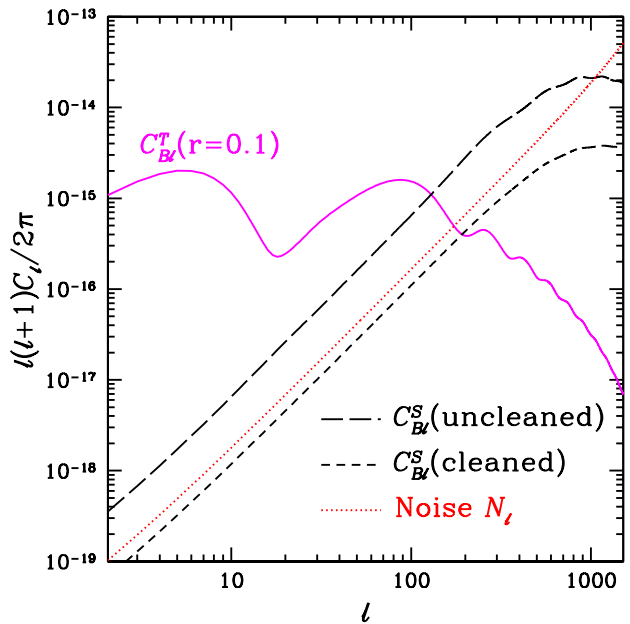}
    \caption{The tensor B mode power spectrum for
$r=0.1$ and $\tau=0.17$ (solid) and the scalar B mode power spectrum 
(long dashes).  The noise power spectrum for an experiment with  
$\Delta_P=3\sqrt{2}\mu K\cdot {\rm arcmin}$ and $\theta_b=3'$ is shown
with the dotted line.  The short-dashed curve is the residual scalar
B mode power spectrum after cleaning of the lensing contaminant by
such an experiment.
}
\end{center}
\end{figure}

\section{Forecasting Detectability Limits}
The cosmic microwave background (CMB) anisotropies are polarized at
the last-scattering surface due to the quadrupolar temperature
fluctuations seeded by the almost scale-invariant density fluctuations at
horizon crossing.  The different types of polarization are expected
from the different sources of the quadrupolar temperature
fluctuations: scalar, vector and tensor quadrupole anisotropy. The
observable polarization patterns are separated into the gradient
E-mode polarization having even parity and the curly B mode
polarization having odd parity \cite{kamionkowski97,seljak97}.

The different sources of the quadrupolar temperature fluctuations
contribute the polarization patterns in their 
own way. 
The scalar source leads to E-mode primarily, but the weak lensing effect
due to the mass distribution along lines of sight between the observer
and the last scattering leads to the transformed secondary B mode \cite{zaldarriaga98}
of which amplitude is naturally smaller 
than the overall amplitude of the primary polarization.
Vector perturbations have no growing modes in linear perturbation theory
and thus are not expected to be significant in inflationary
models for either E or B modes.
The tensor source expected from the inflationary model 
has both E and B patterns in roughly equal magnitude
\cite{kamionkowski97,seljak97}.  Thus the B mode has the
highest ratio of tensor-to-scalar fluctuation power and is
what we consider.

We show the scalar power spectrum $C^S_{Bl}$ and the tensor power
spectrum $C^T_{Bl}$ in Fig. 1 \footnote{We use CMBfast\cite{seljak96} for all
our angular power spectrum calculations.}.  With $r=0.1$, $C^T_{Bl}$ is greater
than $C^S_{Bl}$ at $l<100$.  The bump in $C^T_{Bl}$ at $l<20$ appears
due to reionization.  The amplitude is proportional to the square of
the optical depth, $\tau$ \cite{zaldarriaga97a}; 
here we have set $\tau=0.17$.  
The detectability of the tensor B mode is enhanced by the reionization
bump \cite{knox02,kaplinghat03b}.

The primary temperature and polarization maps are distorted by
the gradient of the projected lensing potential $\phi$.
We can estimate $\phi$ from the 4-point function 
of the temperature and polarization fields \cite{okamoto02}.
With $\phi$ estimated from the lensed maps, we can clean the lensed B mode.
The cleaned and uncleaned $C^S_{Bl}$ are shown in Fig. 1.
The residual lensing-induced B mode after cleaning 
is up to 10 times smaller than the uncleaned lensing-induced B mode.
The minimum detectable limit of $r$ which we can achieve from 
the mass reconstruction is close to $2\times 10^{-5}$ 
which is 10 times better than with no cleaning \cite{knox02,kesden02}.

We use the the following cosmological parameters : 
$\Omega_m=0.34$, $\Omega_b=0.05$, $\Omega_V=0.66$, $h=0.66$, 
$\tau_{reion}=0.17$
and $\sigma_8=0.86$, and let $n_T$ and $r$ be free parameters.
We use the high sensitive future CMB experiments with 
noise levels, $\Delta_P/\sqrt{2}=\Delta_T$ in the range
$\Delta_T=1\mu K\cdot {\rm arcmin}$ to $15\mu K\cdot {\rm arcmin}$,
angular resolution in the range $\theta_b=1.0'$ to $30.0'$
and full sky coverage ($f_{sky}=1$).
Here $\Delta_P=\omega^{-1/2}$ and $\omega$ is the weight per solid angle
for the $Q$ and $U$ linear polarization Stokes parameters.

The variance of B mode CMB power spectrum, $\Delta C^B_{l}$, is
given by
\ba
\Delta C^B_{l}=\sqrt{2 \over (2{l}+1)f_{sky}}
\left(C_{l}^{B,T}+C_{l}^{B,S}+N_{l}
\right),
\ea
where $C_{l}^{B,T}$ is the tensorial B mode and
$C_{l}^{B,S}$ is the scalar B mode
from lensing either before or after cleaning as in \cite{knox02}.
$N_{l}$ is the experimental noise which is given by
\ba
N_{l}=\left({\pi \over 180\times60}\,
{\Delta_P \over {\rm T_{CMB}}}\right)^2
{\rm e}^{l^2 \theta_b^2/(8\ln 2)}.
\ea

Based on the Gaussianity of perturbations,
we use the Fisher matrix analysis to estimate the detectability
\ba
{\rm \bf F}_{pp'}=\sum_{l}
{\partial C^B_{l} \over \partial p}
\left(\Delta C^B_{l}\right)^{-2}
{\partial C^B_{l} \over \partial p'},
\ea
where $p$ and $p'$ enumerate the cosmological parameters we consider here.
The diagonal elements of the ${\rm \bf F}_{pp'}^{-1/2}$
are the 1-$\sigma$ errors for the parameters \cite{press92}.

The slow-roll parameter $\epsilon$ can be replaced by
the measurable quantity $r$, the tensor to scalar ratio $T/S$,
by using the fitting formula \cite{knox95} \cite{turner95},
\ba\label{epr}
\epsilon={1 \over 14}\,
\frac{1.04-0.82\Omega_V+2\Omega_V^2}{1.0-0.03\Omega_V-0.1\Omega_V^2}\,\,r.
\ea
The above fitting formula includes the non-negligible 
contribution to the quadrupole anisotropy from the
late-time ISW effect. 

\begin{figure}[htbp]
\label{f2}
  \begin{center}
    \plotone{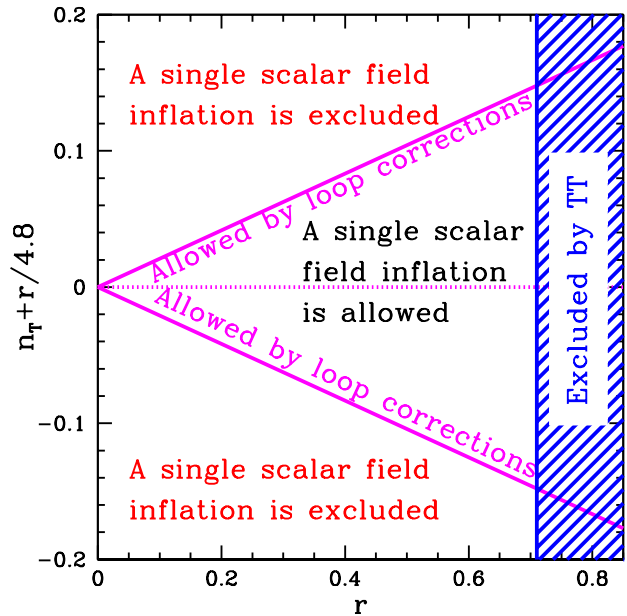}
    \caption{The solid lines bound the possible departures from the consistency
equation for models with a single scalar field.  The shaded area is excluded by observations of the temperature power spectrum \cite{spergel03}.
}
\end{center}
\end{figure}

\section{results and discussion}
Before turning to our detectability limits, we survey the
$r$ vs. $n_T+2(A_T)^2/(A_S)^2$ plane.
We first rewrite the consistency equation Eq.~\ref{consis} with Eq.~\ref{epr},
\ba\label{ntr}
n_T+\frac{1}{4.8}r=-\frac{1}{4.8}\,r\, r_l,
\ea
where $r_l$ is the loop corrections, $r_l\equiv\chi_S H^2/M^2$.
The solid line in Fig. 2 is given by setting $r_l=1$,
its maximum value.
These are largest possible departure to the consistency equation.
If the observed value of $n_T+r/4.8$ is
outside the two solid lines in Fig. 2,
the single scalar field slow-roll
inflation model can be excluded.
If the observed value of $n_T+r/4.8$ is
inside the two solid lines and yet different from zero,
we may have a chance to study physics at distances shorter than $1/M$.
The current upper bound on $r$ from the temperature power spectrum
is 0.71 \cite{bennett03}.

Our detectability limits are shown in Fig. 3.
We fixed all parameters except $r$ and $n_T$ and calculated the Fisher matrix
with $r$ varying from $10^{-5}$ to 0.71.  We show the error on
the combination $n_T + r/4.8$ since we are interested in testing
the consistency equation, but $\sigma(n_T) \simeq \sigma(n_T+r/4.8)$
since $\sigma(n_T) >> \sigma(r)/4.8$.  
As expected, $\sigma(n_T)$ decreases as $r$ increases and raises the
signal level.

\begin{figure}[htbp]
\label{f3}
  \begin{center}
    \plotone{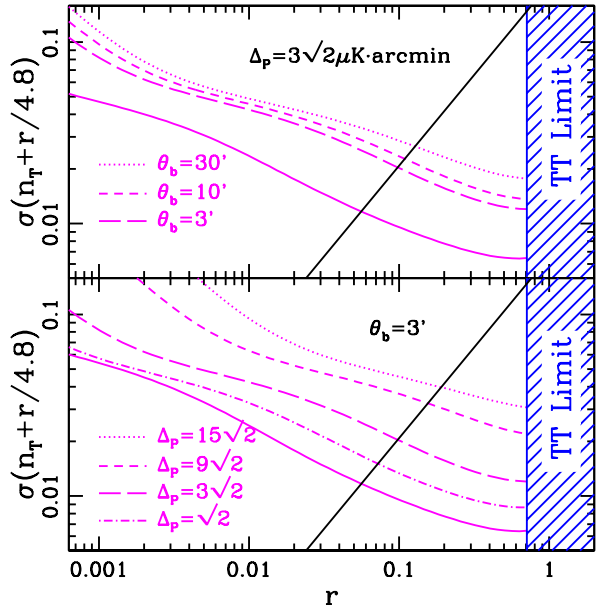}
    \caption{The error on $n_T+r/4.8$ as a function of $r$.  The
cosmic variance limit is the solid line in both panels.  In the upper
panel, we fix $\Delta_P=3\sqrt{2} \mu K\cdot {\rm arcmin}$ and vary
the angular resolution: $\theta_b=1'$ (long dash), $\theta_b=3'$
(dash) and $\theta_b=5'$ (dots).  In the lower panel, we fix the angular
resolution at $\theta_b=3'$ and vary the weight per solid angle:
$\Delta_P=\sqrt{2} \mu K\cdot {\rm arcmin}$
(dot-dashed), $\Delta_P=3\sqrt{2} \mu K\cdot {\rm arcmin}$ (long dash),
$\Delta_P=9\sqrt{2} \mu K\cdot {\rm arcmin}$ (dash) and
$\Delta_P=15\sqrt{2} \mu K\cdot {\rm arcmin}$ (dots).  The shaded area
is excluded by observations of the temperature power spectrum
\cite{spergel03}.  }
\end{center}
\end{figure}

From Fig. 3 we see that at $r<0.1$, any possible theoretical
correction to the consistency equation for slow-roll single field
models is beneath the detectability limit.  Therefore, observing
non-zero $n_T+r/4.8$ and $r < 0.1$ will lead us to exclude the single
slow-roll scalar field inflation scenario.  At $r>0.1$, the loop
correction terms begin to show up above the detectability limit. In
this case, the broken consistency equation does not necessarily mean
the failure of the single slow-roll scalar field inflation.  The solid
straight line indicates the maximal loop corrections with $r_l=1$.
Only the broken consistency above this maximal loop corrections will
mean the failure of the single slow-roll scalar field inflation at
$r>0.1$.

We now consider all the next order corrections in Eq.~\ref{con1}.
The term, $2(A_T^4/A_S^4)=(r/4.8)\times(r/9.6)$, is always below the 
detectable limit with $r<0.71$. 
Also $2(A_T^2/A_S^2)(1-n_S)=r/4.8(1-n_S)$ is well below the 
detectability limit in case of $n_S\sim1$. Even if $n_S$ is much different 
from 1, we can control this term with the knowledge of $n_S$ from
CMB scalar power spectrum. 
The future CMB experiment can determine the $n_S$ within 
$\sigma(n_S)=0.0024$ \cite{kaplinghat03b}.
Thus it is obvious that no other correction terms is larger than
the maximal loop correction terms, i.e. the solid straight line in Fig.~2
is the maximum theoretical bound which the correction terms in
the consistency equation can reach.

In the upper panel of Fig. 3 we show the variation of the angular
resolution with fixed $\Delta_P$.  As the angular resolution
decreases, the ability to clean out the contaminating scalar B mode
diminishes and the detectability limit increases.  We see that at $r >
0.1$ cleaning can make up to a $\sim$50\% decrease in the detectability
limit.  This is due to the improved measurement in the $l=20$ to $200$
range.  At $r \simeq 0.01$ high angular resolution is less important,
since the dominant source of information is now at $l < 20$ which is
unaffected by the scalar contamination.

In the lower panel of Fig. 3 we show the variation of $\Delta_P$
with fixed angular resolution.  As $\Delta_P$ increases, the noise
power becomes larger than the cleaned, and then even the uncleaned,
scalar B mode power.  This increased noise reduces the maximum
observable tensor $l$ and adds significant noise all across the tensor
B mode spectrum rise from $l \simeq 20$ to $l=100$.  
Thus there is a strong sensitivity to increases in 
the noise above our fiducial value.  

The cosmological parameter with the most impact on the tensor spectrum
is $\tau$.  Although $\tau$ can make a big difference for the
detectability limit of $r$ \cite{knox02,kaplinghat03b}, it has little
impact on the detectability limits for $n_T+r/4.8$.  The $r$ limit is
improved by increased $\tau$ since the `reionization bump' at $l < 20$
has $C_l \propto \tau^2$.  In contrast, for acceptable values of
$\tau$, and $r \ga 0.01$, the error in $n_T+r/4.8$ is dominated by
uncertainties in the tensor power spectrum at $l > 20$.  Here the only
effect of reionization is a suppression of power by $exp(-2\tau)$.

Our analysis has ignored polarized emission from galactic and
extragalactic sources.  Multi--frequency observations can
be used to clean out these signals based on their distinct 
spectral shapes.  However, residual contamination is unavoidable
and will also limit the ability of observations to study
the consistency equation.  As we learn more about polarized
foreground emission, these will likely have a big impact
on observing strategies and forecasted $n_T+r/4.8$ detectability limits
below some value of $r$.  Our forecasts should therefore be viewed 
as lower limits. 

\begin{figure}[htbp]
\label{f4}
  \begin{center}
    \plotone{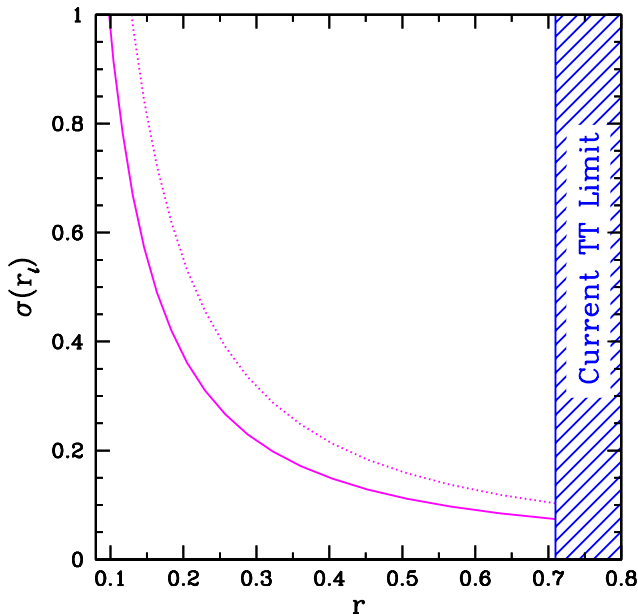}
    \caption{The detectable region of the loop corrections, $r_l$.
The signature of the short distance physics is visible above the curve.
The high resolution reference experiment here is 
$\Delta_P=3\sqrt{2}\mu K\cdot {\rm arcmin}$ $\&$ $\theta_s=3'$. 
The dotted curve is the constraint from the uncleaned B mode.
The solid curve is the constraint from the cleaned B mode.
}
\end{center}
\end{figure}

We now turn our attention to the signature of the short distance physics; i.e.,
how well can we detect a non-zero value of $r_l \equiv \chi H^2/M^2$?
Fig. 4 shows the narrow window for the detectability of $r_l$ 
which can be seen at $r>0.1$.
As Kaloper {\it et. al.} pointed out, we cannot see the new physics
at $r_l< 0.1$, but some M theories with proper compactification
give $r_l> 0.1$ \cite{kaloper02}.

The window for probing the new physics is very narrow at $r>0.1$ and $r_l>0.1$.
But it is a clean window from any next order correction in the slow roll
parameters, since the next order corrections in Eq.~\ref{con1} is
beneath the detectability limit at this window.
We conclude it is possible to probe physics at distances near
$1/M$ as long as $r> 0.1$ and $M$ is very close to $H$.

The dotted line in Fig. 4 is the forecasted constraint from the
uncleaned B mode with the same experiment ($\Delta_P=3\sqrt{2}\mu
K\cdot {\rm arcmin}$ $\&$ $\theta_b=3'$).  We get the solid line by
cleaning the scalar B mode with the estimated lensing potential.  If
the reduced Planck mass is truly such a small amount, i.e. $r_l>0.1$,
then even a small amount of improvement in $\sigma(r_l)$ will be
valuable.  As we see in Fig. 4, the lensing potential reconstruction 
improves the detectability of $r_l$ by about 50\%.
If we can detect the tensor power spectrum at $r>0.1$, the uncleaned B
mode can probe the tensor power spectrum well. But we will want 
the cleaned B mode to probe to even shorter distances.

Wands et al. \cite{wands02} have studied the consistency equation
for two and more scalar fields.  For two scalar fields there is
a generalized consistency equation giving the tensor perturbation to
scalar curvature perturbation ratio proportional to $n_T$ times
an isocurvature correlation angle.  For more than two fields this
becomes an inequality, providing an upper bound on this tensor-to-scalar
ratio.  Thus observation of departure from this generalized consistency
equation (by more than allowed by loop corrections) would rule
out two--field models and an observed violation of the inequality
would rule out {\em all} slow-roll models of inflation.

\section{Conclusions} 

The tensor perturbation spectrum is a more direct probe of inflation
than the scalar perturbation spectrum and may provide us with highly
valuable information about the physics of inflation.  We have
quantified how well idealized versions of future experiments can probe
the tensor perturbation spectrum and, in particular, test the
inflationary consistency equation.  Detectable departures may
come from additional fields or short--distance corrections to the
inflaton's effective field theory.  We have shown that for $r \ga
0.01$, even though high resolution does not improve the detectability
of $r$, high resolution does improve the detectability of departures
from the consistency equation.

We thank Nemanja Kaloper for his feedback on a 
draft version of this manuscript and Andreas Albrecht, Eugene Lim and 
Sayan Basu for useful conversations.  This work was supported
in part by NASA NAG5-11098.

\bibliography{cmb3}
\end{document}